\documentclass[preprint,12pt]{elsarticle}
\newcommand{\beqn}{\begin{equation}}
\newcommand{\eeqn}{\end{equation}}
\newcommand{\ba}{\begin{eqnarray}}
\newcommand{\ea}{\end{eqnarray}}

\newcommand{\vk}{{\bf k}}

\newcommand{\vB}{{\bf B}}

\newcommand{\sLH}{\mbox{\tiny LH}}
\newcommand{\sUH}{\mbox{\tiny UH}}

\def\la{\mathrel{\hbox{\rlap{\hbox{\lower4pt\hbox{$\sim$}}}\hbox{$<$}}}}
\def\ga{\mathrel{\hbox{\rlap{\hbox{\lower4pt\hbox{$\sim$}}}\hbox{$>$}}}}
\def\df{\mathrel{\hbox{\rlap{\hbox{\lower-6pt\hbox{$\small{df}$}}}\hbox{$=$}}}}

\usepackage{amssymb}

\journal{Fundamental Plasma Physics}

\begin{document}

\begin{frontmatter}

\title{Model expressions for refractive indices of electron waves
in cold magnetoactive plasma of arbitrary density}
\author{D.R. Shklyar$^{*,}$\footnote{Space Research Institute of RAS (IKI),
Profsoyuznaya str. 84/32, 117997, Moscow, Russia.
Tel.: +(7-495) 333 45 34 Fax: +(7-495) 333 33 11
\mbox{E-mail:} david@iki.rssi.ru}, N.S. Artekha$^{*,**}$}

\address{$^*$Space Research Institute of RAS, Moscow, Russia}
\address{$^{**}$HSE University, Moscow, Russia}

\begin{abstract}
Despite the undoubted importance of having fairly simple analytical expressions for the refractive indices
of wave modes in a magnetoactive plasma, such expressions are known only in some particular cases.
For electron waves with frequencies much higher than the lower hybrid resonance frequency,
such an expression is known only for whistler waves in a dense plasma when the electron plasma frequency
significantly exceeds the electron cyclotron frequency. In this Letter, we propose simple operational
expressions for the refractive indices of all four electron modes in a magnetoactive plasma, namely,
the fast magnetosonic, also called whistler mode, the slow extraordinary mode, the ordinary mode,
and the fast extraordinary mode. The form of these expressions does not depend on the value of the ratio
of plasma frequency to cyclotron frequency.
\end{abstract}

\end{frontmatter}
Wave refractive index is a fundamental characteristic of the wave mode in plasma 
\cite{Akhiezer,Bittencourt,Ginzburg_astro,Krall,Shafranov}.
It determines the wave propagation features, the wave polarization, as well as the wave energy density.
In a cold magnetoactive plasma it is determined by the biquadratic equation
\beqn
A N^4 + B N^2 + C = 0 \; ,
\label{N^4}
\eeqn
which is derived and discussed in most textbooks on plasma physics
(see, e.g., \cite{Akhiezer,Ginzburg, Lifshitz_Pitaevskii,Shafranov,Stix}).
Here $N = kc/\omega$ is the wave refractive index, $k$ is the magnitude of the wave normal vector,
 $\omega$ is the wave frequency, and $c$ is the speed of light. The coefficients of equation (\ref{N^4})
are expressed through components of dielectric tensor
\beqn
\varepsilon_{ij}(\omega) = \left(\begin{array}{ccc}
\varepsilon_1 & i \varepsilon_2 & 0 \\
- i \varepsilon_2 & \varepsilon_1 & 0 \\
0 & 0 & \varepsilon_3
\end{array}
\right )
\label{dielectric_tensor(omega)}
\eeqn
and the wave normal angle $\theta$, i.e., the angle between the wave vector $\vk$ and the ambient
magnetic field $\vB_0$, directed along the $z$-axis, in the following way
\ba
\label{A,B,C}
A &=& \varepsilon_1 \mbox{sin}^2\theta + \varepsilon_3 \mbox{cos}^2\theta \;
\nonumber \\
B &=& - \varepsilon_1 \varepsilon_3 (1 + \mbox{cos}^2 \theta)
- (\varepsilon_1^2 - \varepsilon_2^2) \mbox{sin}^2\theta \; ;\\
C &=& \varepsilon_3 (\varepsilon_1^2 - \varepsilon_2^2) \; ; \nonumber
\ea
For electron waves with frequencies $\omega$ much larger than lower hybrid resonance frequency
$\omega_{\sLH}$, the real quantities $\varepsilon_1 , \; \varepsilon_2 , \;$ and
$\varepsilon_3$ are given by
\beqn
\varepsilon_1 = \frac{\omega^2 - \omega_{\sUH}^2}{\omega^2 - \omega_c^2} \; ; \; \; \;
\varepsilon_2 = \frac{\omega_p^2 \omega_c}{\omega(\omega^2 - \omega_c^2)}  \; ; \; \; \;
\varepsilon_3 = \frac{\omega^2 - \omega_p^2}{\omega^2} \; ,
\label{epsilons}
\eeqn
where $\omega_p$ and $\omega_c$ are electron plasma frequency and the magnitude of electron
cyclotron frequency, respectively, and $\omega_{\sUH} =\sqrt{\omega_p^2 + \omega_c^2}$
is the upper hybrid frequency. Solutions of the equation (\ref{N^4})
\beqn
N_{1, 2}^2 = \frac{- B \pm \sqrt{B^2 - 4 A C}}{2 A} \; .
\label{N^2}
\eeqn
give two possible values of refractive index squared for given frequency and wave normal angle,
which also depend on $\omega_p$ and $\omega_c$, of course. For given values of these parameters,
both quantities $N_{1, 2}^2$ may be positive determining two propagating modes, one $N^2$
may be positive and one negative, giving one propagating mode, or both quantities  $N_{1, 2}^2$
may be negative, meaning that for these frequency, wave normal angle, and plasma parameters
there are no propagating electron modes in the approximation of cold plasma. As the analysis shows,
the parameter that largely determines the properties of waves is the ratio
of the plasma to cyclotron frequency, which we will denote by $p$:
\beqn
p = \frac{\omega_p}{\omega_c} \; .
\label{p}
\eeqn
As is well known, in cold magnetoactive plasma there are four electron modes.
According to the denomination used by Shafranov \cite{Shafranov}  
and Akhiezer et al. \cite{Akhiezer}, 
they are called fast magnetosonic
(FM or whistler) mode, slow extraordinary (SE) mode, ordinary (O) mode, and fast extraordinary (FE)
mode. In constructing model expressions for refractive indices of electron waves, we will rely upon
the following well known results. In the case of parallel propagation, i.e., $\theta = 0$,
the index of refraction is determined by
\beqn
N^2 \equiv \varepsilon_1 - \varepsilon_2 = \frac{\omega^2 - \omega\omega_c - \omega_p^2}{\omega(\omega - \omega_c)} \; ,
\label{WandX}
\eeqn
or by
\beqn
N^2 \equiv \varepsilon_1 + \varepsilon_2 = \frac{\omega^2 + \omega\omega_c - \omega_p^2}{\omega(\omega + \omega_c)}\; .
\label{SEandO}
\eeqn
For $\theta = 0$, there also exists electrostatic Langmuir wave at the frequency $\omega = \omega_p$
for which the refractive index may be arbitrary. We should mention that the case of parallel propagation
is degenerated if the frequency band of the corresponding mode includes $\omega = \omega_p$,
since in this case all the coefficients of the equation (\ref{N^4}) turns to zero, and the refractive index
is not well defined. That is why the refractive index of the same wave mode may be described by different
expressions (\ref{WandX}) or (\ref{SEandO}) in the frequency domains below and above $\omega_p$,
and this change occurs abruptly. This does not happen for whistler mode if $p > 1$, since in this case
$\omega<\omega_p$, and for FE mode whose frequency is always larger than $\omega_p$.
In both of these cases, the expression (\ref{WandX}) is in effect.

In the case of perpendicular propagation, i.e., $\theta = \pi/2$, the refractive index is given by
\beqn
N^2 \equiv \frac{\varepsilon_1^2 - \varepsilon_2^2}{\varepsilon_1} =
\frac{(\omega^2 - \omega\omega_c - \omega_p^2)(\omega^2 + \omega\omega_c - \omega_p^2)}
{\omega^2(\omega^2 - \omega_{\sUH}^2)}
\label{SEandFE}
\eeqn
for SE- and FE-modes in the corresponding frequency bands, and by
\beqn
N^2 \equiv \varepsilon_3 = \frac{\omega^2  - \omega_p^2}{\omega^2}
\label{O}
\eeqn
for O-mode.

An important characteristic of the wave mode, if exists, is the cutoff frequency, determined from
$C = 0$, at which the refractive index equals to zero. These cutoff frequencies are given by
\beqn
\omega_{01} = \frac{\omega_c}{2} + \sqrt{ \frac{\omega_c^2}{4} + \omega_p^2} \; ; \; \; \;
\omega_{02} = \omega_p \; ; \; \; \;
\omega_{03} = - \frac{\omega_c}{2} + \sqrt{ \frac{\omega_c^2}{4} + \omega_p^2}
\label{cutoffs}
\eeqn
for FE-, O-,  and SE-mode, respectively. Other characteristic frequency of the wave mode is the resonance
frequency, determined from $A = 0$, at which the refractive index tends to infinity.
In the domain of electron wave there are two of them:
\beqn
\omega_{r1}^2 = \frac{\omega_{\sUH}^2}{2} +
\sqrt{ \frac{\omega_{\sUH}^4}{4} - \omega_p^2 \omega_c^2 \cos^2 \theta}  \; ; \; \; \;
\omega_{r2}^2 = \frac{\omega_{\sUH}^2}{2} -
\sqrt{ \frac{\omega_{\sUH}^4}{4} - \omega_p^2 \omega_c^2 \cos^2 \theta}  \; ; \; \; \;
\label{r1,2}
\eeqn
which correspond to SE- and whistler-mode, respectively. We should mention that the cutoff frequencies
do not depend on propagation angle, while the resonance frequencies do. The latter have the following
limit values
\ba
\label{limits}
\omega_{r1}(\theta = 0) = \max(\omega_p, \omega_c) \; ; \; \;
\omega_{r1}(\theta = \pi/2) = \omega_{\sUH} \; ; \nonumber\\
\omega_{r2}(\theta = 0) = \min(\omega_p, \omega_c) \; ; \; \;
\omega_{r2}(\theta = \pi/2) \to 0 \; ; \\
\omega_{r2}(p \gg 1) =  \omega_c |\cos \theta| \; ; \; \; 
\omega_{r2}(p \ll 1) =  \omega_p |\cos \theta| \; . \nonumber
\ea
Note that when ion contribution to the dielectric tensor is neglected, whistler waves cannot propagate
at $\theta \to \pi/2$ since the range of permitted frequencies inside the resonance cone vanishes.
The results given above can be found in many textbooks on plasma physics
\cite{Akhiezer,Shafranov,Stix}, as well as in others cited before,
and serve as a basis for the following development.

Despite the importance of the wave refractive index for plasma physics,
its explicit expression as a function of frequency and wave normal angle, which would not contain radicals,
is rarely known. Regarding electron waves, it is known only for whistler mode waves in the case of dense
plasma $p \gg 1$. This expression may be obtained in the following phenomenological way.
For $p \gg 1$ and $\omega < \omega_c$ the expression (\ref{WandX}) turns into
\beqn
N^2  = \frac{\omega_p^2}{\omega(\omega_c - \omega)} \; .
\label{W}
\eeqn
Since for $p \gg 1$, $\omega_c$ is the resonance frequency of whistler waves at $\theta=0$,
for arbitrary angle of propagation it is natural to replace $\omega_c$ by $\omega_{r2}$
(\ref{r1,2}), which for $p \gg 1$ gives $\omega_c |\cos \theta|$. This gives the well-known  expressions
for the refractive index and the dispersion relation  of whistler waves:
\beqn
N^2  = \frac{\omega_p^2}{\omega(\omega_c |\cos \theta|  - \omega)} \; ;  \; \; \;
\omega = \omega_c |\cos \theta| \frac{k^2 c^2}{k^2 c^2 + \omega_p^2} \; .
\label{N_and_frequency}
\eeqn
The validity of these expressions can then be checked by direct calculations using (\ref{N^4})
and the condition $p \gg 1$. In this Letter, we obtain, in the same spirit as above, approximate expressions
for the refractive indices of four electron modes in wide ranges of wave normal angles
and arbitrary ratio $p$ between electron plasma and cyclotron frequencies. Since the corresponding
analytical expressions do not exist, we check the accuracy of the obtained expressions against exact solutions
of the equation (\ref{N^4}).

For whistler waves, the model expression for $N^2$ has the form:
\beqn
N^2  = \frac{(\omega_p^2 + \omega \omega_c - \omega^2)[\min(\omega_c, \omega_p) - \omega]}{\omega(\omega_c - \omega)(\omega_{r2} - \omega)} \; ,
\label{W_general}
\eeqn
where $\omega_{r2}$ is defined in (\ref{r1,2}). This expression gives the exact value of $N^2$
at $\theta = 0$ and all $\omega$ in the whistler mode band, the exact resonance frequency for all $\theta$
and $p$, and goes into (\ref{N_and_frequency}) for $p \gg 1$. Indeed, in this case the last two terms 
in the first bracket in the numerator may be neglected, the second bracket equals to $(\omega_c - \omega)$
and is canceled by the first one in the denominator, while $\omega_{r2} = \omega_c |\cos \theta|$
(see (\ref{limits})), and we come to the expression for $N^2$ in (\ref{N_and_frequency}).
In the case $p \ll 1$, the first bracket in the numerator cannot be simplified, 
the second one gives $(\omega_p - \omega)$,  $\omega_{r2} = \omega_p |\cos \theta|$ (see (\ref{limits})),
and we obtain
\beqn
N^2  = \frac{(\omega_p^2 + \omega \omega_c - \omega^2)(\omega_p - \omega)}
{\omega(\omega_c - \omega)(\omega_p |\cos \theta| - \omega)} \simeq
\frac{(\omega_p^2 + \omega \omega_c - \omega^2)(\omega_p - \omega)}
{\omega \omega_c (\omega_p |\cos \theta| - \omega)} \; ,
\label{W_small_p}
\eeqn
which gives especially simple result in the case of parallel propagation:
$$N^2  = \frac{\omega_p^2 + \omega \omega_c - \omega^2}{\omega \omega_c} \; .$$

Unlike whistler waves, expressions for $N^2$ of the remaining three modes at arbitrary
propagation angles $\theta$ will be based on  the exact expressions
(\ref{SEandFE}) and (\ref{O}) for perpendicular propagation. We begin with FE mode.
Expression (\ref{SEandFE}) gives exact value of $N^2$ at $\theta = \pi/2$ for all frequency,
being positive for $\omega$ above the cutoff frequency $\omega_{01}$ and negative below.
Since for FE wave $N^2$ goes to $- \infty$ when $\omega \to \omega_{r1}+0$,
i.e., at the resonance frequency of SE wave, and since $\omega_{r1}(\pi/2) = \omega_{\sUH}$,
a natural generalization of the formula (\ref{SEandFE})  to FE wave is
\beqn
N^2  = \frac{(\omega^2 - \omega\omega_c - \omega_p^2)(\omega^2 + \omega\omega_c - \omega_p^2)}
{\omega^2(\omega^2 - \omega_{r1}^2)} \; .
\label{FE}
\eeqn
Expression (\ref{FE}) is exact for $\theta = \pi/2$, gives correct cutoff frequency which is independent
of $\theta$, as it should, gives correct $\theta$-dependent frequency at which $N^2 \to - \infty$,
and gives correct behavior of $N^2$ at $\omega \to \infty$, namely, $N^2 \to 1$.
All these properties are fulfilled for arbitrary values of the parameter $p$, leaving little freedom
for a modification of the suggested expression.

Generalization of the expression (\ref{O}) for O mode to arbitrary angles of propagation is similar to the case
of FE mode, with the only difference that the frequency at which the refractive index
should tend to $- \infty$ is now equal to $\omega_{r2}$.
Considering that in the accepted approximation $\omega_{r2}(\pi/2) = 0$ we obtain for O mode
\beqn
N^2 = \frac{\omega^2  - \omega_p^2}{\omega^2 - \omega_{r2}^2} \; .
\label{O_model}
\eeqn
All properties of the model expression for FE mode mentioned above also apply to O mode.

Similar to FE mode, in the case of SE mode we  proceed from (\ref{SEandFE}) and write
the expression for $N^2$ in the form
\beqn
N^2 =
\frac{(\omega^2 - \omega\omega_c - \omega_p^2)(\omega^2 + \omega\omega_c - \omega_p^2)}
{\omega^2(\omega - \omega_{r1})(\omega + \omega_r)} \; ,
\label{SE}
\eeqn
where $\omega_{r1}$ is defined in (\ref{r1,2}), and $\omega_r$ remains to be determined.
Expression (\ref{SE}) gives correct cutoff frequency, and correct resonance frequency $\omega_{r1}$
for SE mode. Since $\omega_{r1}(\theta = \pi/2) = \omega_{\sUH}$, the expression (\ref{SE})
would give exact values of $N^2$ at $\theta = \pi/2$ for all frequencies, provided that
$\omega_r (\theta = \pi/2)$ is also equal to $\omega_{\sUH}$. To determine the quantity
$\omega_r$, we will use a general property of the refractive index for the SE wave,
which apparently has not been stated before, namely, that at $\omega=\omega_p$, $N^2 = 1$
for all values of parameters and all wave normal angles. Indeed, at $\omega = \omega_p$,
$A = \varepsilon_1 \sin^2\theta$, $B = - ( \varepsilon_1^2 - \varepsilon_2^2)\sin^2\theta$, $C = 0$,
and from general equation (\ref{N^4}) we have $N^2 = (\varepsilon_1^2 - \varepsilon_2^2)/\varepsilon_1$.
Thus, $N^2(\omega = \omega_p)$ does not depend on $\theta$ and coincides with the expression for
$\theta = \pi/2$:
\beqn
N^2 = \frac{\omega_p^2 \omega_c^2}{\omega_p^2(\omega_{\sUH}^2 - \omega_p^2)} = 1 \; .
\label{N^2=1}
\eeqn
From the condition that at $\omega = \omega_p$ the expression (\ref{SE}) equals $1$
for all wave normal angles $\theta$, we find
\beqn
\omega_r =
\frac{\omega_{\sUH}^2 - \omega_p \omega_{r1}}{\omega_{r1} - \omega_p} \;
\label{omega_r}
\eeqn
Using the definition (\ref{r1,2}) one can easily check that $\omega_r (\theta = \pi/2) = \omega_{\sUH}$.
A comparison between exact and model expressions $N^2(\omega)$ of electron modes
for four values of the wave normal angle and three values of the parameter $p$
is given in Figure 1.

In sum, we have presented model expressions for the wave refractive indices
of electron waves in cold magnetoactive plasma that give good approximations 
to exact values in a wide range of parameters: wave frequencies, wave normal angles, and the ratios 
of electron plasma frequency to electron cyclotron frequency. 
A natural question to arise is about the applicability of the model expressions for $N^2$
in different domains of parameters.
In this Letter, we will limit ourselves to a qualitative discussion of this matter. 
As one can see from the Figure 1, the approximate expression for whistler waves works well 
for $p \gg 1$ and $p \ll 1$ for all wave normal angles $\theta$, and gives quite satisfactory
agreement with exact values for $p \simeq 1$. 
Model expression for SE mode works quite well for most angles $\theta$  and values 
of the parameter $p$, except small $\theta$ and large $p$. The expression for $O$-mode 
doesn't work well for small $\theta$ and $p \la 1$, while the expression for FE mode 
works well for all values of $\theta$ and $p$, except small $\theta$ and $p \sim 1$. 
The fact that for the last three modes the model expressions fail at small $\theta$ 
is not surprising since in their derivations we proceeded from the rigorous expressions 
for $\theta = \pi/2$ at which our model expressions are exact.\\


\begin{figure}
\centering
\vspace*{-5cm}
\hspace*{-2cm}
\includegraphics[scale=.75]{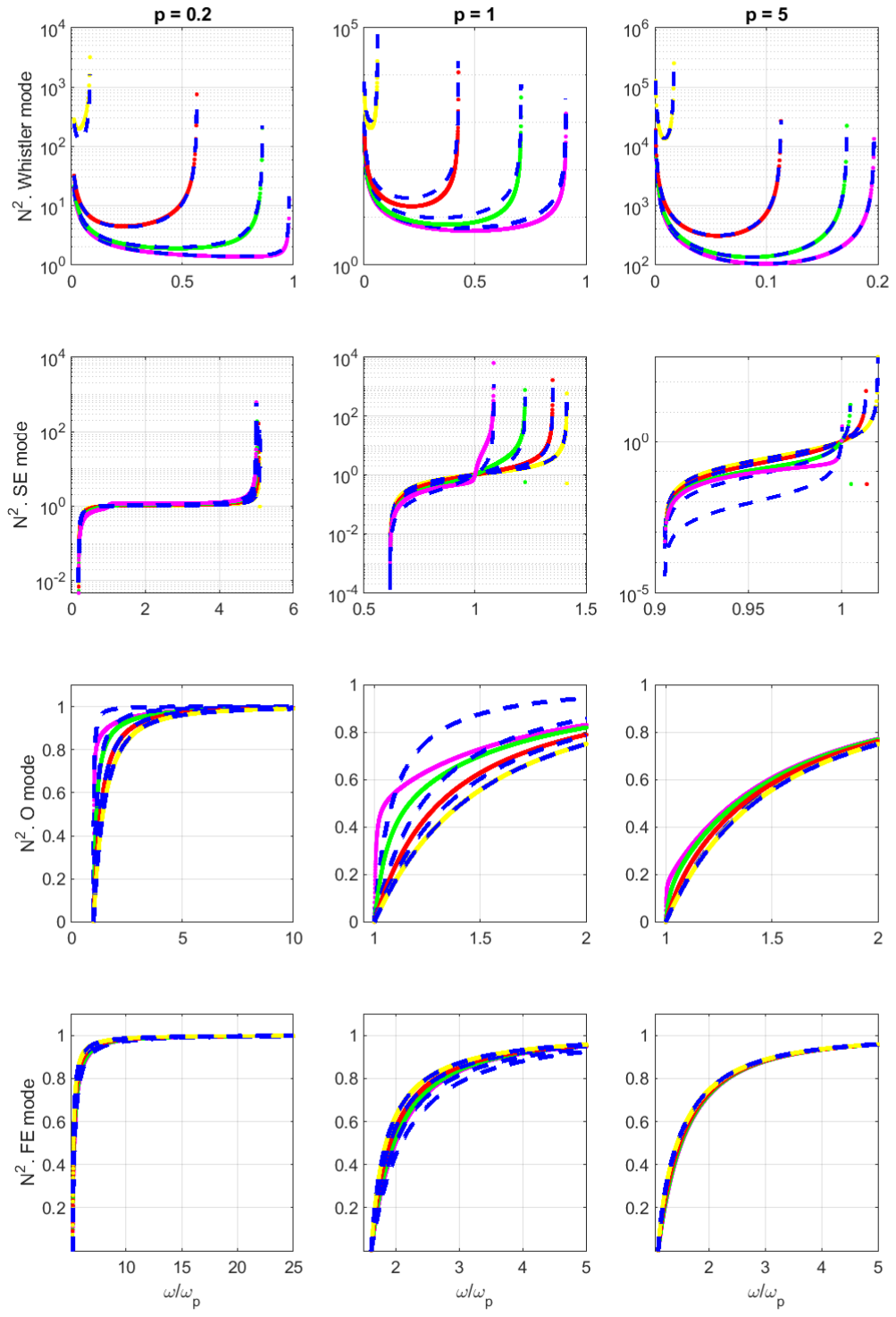}
\vspace*{-1cm}
\caption{Refractive indices squared as functions of normalized frequency. 
Exact values for four wave normal angles $10^{\circ}, 30^{\circ}, 55^{\circ}, \mbox{and} \; 85^{\circ}$
are shown in magenta, green, red, and yellow, respectively, and model quantities are shown in blue.
The quantity $N^2(\omega)$ is drawn for four electron modes: whistlers, SE, O, and FE-mode 
from top to bottom row, respectively, and for three values of the parameter $p=\omega_p/\omega_c$ 
indicated above the corresponding columns.}
\label{fig_nsqes}
\end{figure}
\newpage
\bibliographystyle{plain}
\bibliography{Refractive_indices}

\end{document}